\documentclass[aps,prl,showpacs,twocolumn]{revtex4}
\usepackage{amsmath, amssymb}
\usepackage{graphicx}
\usepackage{subfigure}

\renewcommand{\vec}[1]{\mathbf{#1}}

\DeclareMathOperator{\tr}{Tr}
\DeclareMathOperator{\nthM}{nthMultiindex}
\DeclareMathOperator{\nextM}{nextMultiindex}

\begin{document}

\title{Multipartite continuous-variable entanglement}

\author{E. Shchukin}
\email{evgeny.shchukin@uni-rostock.de}
\author{W. Vogel}
\email{werner.vogel@uni-rostock.de} \affiliation{Arbeitsgruppe Quantenoptik, Institut f\"ur Physik,
Universit\"at Rostock, D-18051 Rostock, Germany}

\begin{abstract}
  Necessary and sufficient observable conditions for the nonnegativity of all
  partial transpositions of multi-mode quantum states are derived. The
  result is a hierarchy of inequalities for minors in terms of moments of
  the given state. Violations of any inequality is a sufficient
  condition for entanglement. Full entanglement can be certified for a
  manifold of multi-mode quantum states. A \textit{Mathematica} package is
  given for a systematic test of the hierarchy of conditions.
\end{abstract}

\pacs{03.67.Mn, 03.65.Ud, 42.50.Dv}

\maketitle

Quantum information is a rapidly developing field of modern physics.
Entanglement is considered to be the key resource for a variety of
applications, such as quantum computation, quantum state teleportation and
others. Recently increasing interest arose in the use of continuous variable
(cv) entangled states in quantum information processing, for a review
see~\cite{rmp-77-513}. So far many such considerations deal with Gaussian
entangled states, due to both the simplicity of characterizing their
entanglement properties and the possibilities of their experimental
realization.

Criteria of bipartite entanglement of Gaussian states are well established.
Even necessary and sufficient conditions for the entanglement of such states
are known and some generalizations have been formulated~\cite{prl-84-2726}.
In the more general case of non-Gaussian quantum states a complete
characterization of entanglement is still an open problem. A powerfull
approach of characterizing an important class of entangled states is based on
the Peres-Horodecki criterion which relies on the positive but not
completely positive map of partial transposition~\cite{prl-77-1413}.
Whenever the nonpositivity of partial transposition (NPT) is identified, the
state is entangled.

Recently the necessary and sufficient conditions for the NPT of a bipartite
quantum state have been formulated by using moments of arbitrarily high
orders~\cite{prl-95-230502}. This approach yields a manifold of new
entanglement conditions and unifies a series of previously known ones,
including such whose derivation was not directly based on NPT. It is also of great importance that an efficient method has been developed for measuring the quantum correlation functions that are needed in these
criteria~\cite{prl-96-200403}. First attempts have been made to extend this approach, to describe bipartite entanglement of more than two modes and entanglement beyond NPT~\cite{prl-96-050503,quant-ph/0605001}.

The characterization of multipartite entanglement is clearly much more
sophisticated than is the bipartite case. For example, multipartite
entanglement is considered to be important for quantum computation. For some
discussion of multipartite entanglement in finite dimensional Hilbert
spaces, cf. e.g.~\cite{quant-ph/0505149}. Furthermore, multipartite
cv entanglement has been studied by using moments of second
order~\cite{njp8-51}. Presently general criteria for multipartite cv
entanglement of arbitrary quantum states are unknown.

In the present contribution we derive necessary and sufficient conditions for
the positivity of all partial transpositions of a general cv multi-mode
quantum state. Hence we obtain a hierarchy of necessary conditions for
separability. The violation of any such condition is sufficient
for entanglement. We show that full multipartite entanglement of a
variety of quantum states can be proven by analyzing all possible partial
transpositions. A \textit{Mathematica} package for completely testing the partial transpositions is given, which can use the measured or calculated quantum correlations of a given state.

There are several notions of separability. Let us start to consider full
separability \cite{pra-40-4277}. A general $n$-mode state $\hat{\varrho}$ is
called fully separable if it is a convex combination of separable states,
\begin{equation}\label{eq:rho}
    \hat{\varrho} = \sum^{+\infty}_{k = 1} p_k \hat{\varrho}^{(1)}_k \otimes \ldots \otimes \hat{\varrho}^{(n)}_k,
\end{equation}
where $\hat{\varrho}^{(i)}_k$ is a state of the $i$th mode ($i = 1, \ldots
n$). The numbers $p_k$ satisfy the conditions $p_k \geqslant 0$,
$\sum^{+\infty}_{k=1} p_k = 1$, and the series converges in the trace-norm.

This definition can be extended in the following natural manner. Let $\pi =
\{I_1, \ldots, I_p\}$ be a decomposition of the set $\mathcal{N}_n$ into a
disjoint union of the subsets $I_i \subseteq \mathcal{N}_n$, $i = 1, \ldots,
p$:
\begin{equation}\label{eq:parsep}
    \mathcal{N}_n = I_1 \cup \ldots \cup I_p, \quad I_i \cap I_j = \varnothing \quad \text{if}\quad i \not= j.
\end{equation}
The number of parts of the decomposition $\pi$ we denote as $|\pi|=p$. The state $\hat{\varrho}$ is
called $\pi$-separable if it can be represented as
\begin{equation}
    \hat{\varrho} = \sum^{+\infty}_{k = 1} p_k \hat{\varrho}^{(I_1)}_k \otimes \ldots \otimes \hat{\varrho}^{(I_p)}_k,
\end{equation}
where $\hat{\varrho}^{(I_i)}_k$ is a state of the multi-mode part of the original system formed by the
modes with the indices in $I_i$, $i = 1, \ldots, p$, and the numbers $p_k$ satisfy the same conditions as in the case of full separability. For a related classification of multipartite qubit states we refer the reader to \cite{pra-61-042314}.

A decomposition $\sigma = \{J_1, \ldots, J_q\}$ is said to be finer than $\pi$, $\sigma \prec \pi$,
if for any $j = 1, \ldots, q$ there is an $i = 1, \ldots, p$ such that $J_j \subseteq I_i$. The
finest decomposition is $\tau = \{\{1\}, \ldots, \{n\}\}$, which corresponds to the notion of full
separability. It is clear that a $\pi$-separable quantum state is also $\sigma$-separable for all
$\sigma$ such that $\pi \prec \sigma$. In particular, a fully separable state is $\pi$-separable
for any $\pi$. Thus it makes sense to consider only minimal elements of the set
$\mathcal{D}(\hat{\varrho})$ of decompositions $\pi$ such that $\hat{\varrho}$ is $\pi$-separable.
In \cite{prl-82-5385} it was shown that there are states $\hat{\varrho}$ such that
$\mathcal{D}(\hat{\varrho})$ has different minimal elements or, in other words, that some states
can be separated in different incomparable ways. The hierarchy of the notions of separability considered here can be ordered as:
\begin{equation}
\begin{split}
\label{eq:sep-order}
    &\text{full separability}\ \to\ \sigma-\text{separability} \\
    &\to\ \pi-\text{separability with}\ \sigma \prec \pi.
\end{split}
\end{equation}
The full separability is the most restrictive one (the smallest number of states are fully separable). Conversely, for entanglement we have: the class of entangled states that violate full separability is the largest, but for many applications not the most useful one.

To determine whether a given multi-mode state is entangled or not is in
general a highly nontrivial problem. A partial and powerful approach, which
can verify entanglement but cannot verify separability, is based on the notion
of partial transposition. The transposition $\mathrm{T}$ of a single-mode
system is defined via $\mathrm{T}(|n\rangle\langle m|) = |m\rangle\langle n|$,
where $\{|n\rangle\}$ is the Fock basis in the Hilbert space associated with
the system. For a multi-mode system one can introduce the notion of partial
transposition, when only some modes are transposed and the others are left
unchanged.

For an $n$-mode system one can construct $2^n$ partial transpositions, which
are in one-to-one correspondence with subsets $I$ of the set $\mathcal{N}_n =
\{1, \ldots, n\}$. The partial transposition corresponding to the subset $I
\subseteq \mathcal{N}_n$ we denote as $\mathrm{PT}_I$. For $I = \varnothing$
and $I = \mathcal{N}_n$ we have $\mathrm{PT}_{\varnothing} = 1$ and
$\mathrm{PT}_{\mathcal{N}_n} = \mathrm{T}_{\mathrm{tot}}$ respectively, where
$\mathrm{T}_{\mathrm{tot}}$ is the total transposition of the multi-partite
system. Since $\mathrm{T}^2 = 1$ we have the relation $\mathrm{PT}_I \circ
\mathrm{PT}_J = \mathrm{PT}_{I \Delta J}$, valid for all $I, J \subseteq
\mathcal{N}_n$, with $I \Delta J = (I \setminus J) \cup (J \setminus I)$ being
the symmetric difference of $I$ and $J$. Taking $J = \mathcal{N}_n$ we get
\begin{equation}\label{eq:complement}
    \mathrm{PT}_{\overline{I}} = \mathrm{PT}_I \circ \mathrm{T}_{\mathrm{tot}},
\end{equation}
where $\overline{I} = \mathcal{N}_n \setminus I$ is the complement of $I$.

One can easily see that the partial transposition of a separable quantum state is again a separable
state, hence it is nonnegative defined.  One can use this fact as a test for entanglement: any
negativity of any partial transposition is a clear signature of entanglement. Note that the trivial
partial transposition $\mathrm{PT}_{\varnothing} = 1$ and the total transposition
$\mathrm{PT}_{\mathcal{N}_n} = \mathrm{T}_{\mathrm{tot}}$ of any state are always quantum states,
so there are $2^n - 2$ choices of the subsets $I \subseteq \mathcal{N}_n$ which can lead to
negativity. Due to the relation \eqref{eq:complement} and the fact that the total transposition of
any system is always nonnegative it is clear that for any subset $I \subset \mathcal{N}_n$ it makes
sense to test only one of the partial transpositions $\mathrm{PT}_I$ and
$\mathrm{PT}_{\overline{I}}$, the other gives the same result. Finally, there are $2^{n-1}-1$
partial transpositions to test, which can be identified with non-empty subsets of
$\mathcal{N}_{n-1}$ (including the set $\mathcal{N}_{n-1}$ itself).

It is not trivial that the study of partial transpositions gives insight in
multipartite entanglement properties.  Partial transposition tests only for
bipartite entanglement since there are only two groups of modes: those we
transpose and those we do not. Nevertheless, our approach allows one to
certify full multipartite entanglement of a given $n$-mode quantum state.
Contrary to the notion of full separability, this requires that the state
under study does not show $\pi$-separability for any decomposition $\pi$. This
may be of some practical interest. In applications of multimode entangled
quantum states the user may need to share $n$ modes among $k$ parties ($k
\leqslant n$) with the requirement of full entanglement.

To certify full entanglement of some $n$-mode cv quantum states, it is
sufficient to demonstrate that all $2^{n-1}-1$ nontrivial partial
transposition are nonpositive.  In fact, if the state is not fully entangled
then there exists a decomposition $\pi$ such that $\hat{\varrho}$ is
$\pi$-separable. According to relation~\eqref{eq:sep-order} it is clear that
any $\pi$-separable state must fulfill at least one bipartite separability
condition.  Hence the proof that all partial transpositions are nonpositive is
a certificate for full entanglement. The details of the practical realization
of such a test will be given below.

A Hermitian operator $\hat{A}$ is nonnegative if and only if $\langle
\hat{f}^\dagger \hat{f} \rangle_A \equiv \tr(\hat{A} \hat{f}^\dagger \hat{f})
\geqslant 0$, for all operators $\hat{f}$ whose normally ordered form exists
\cite{prl-95-230502}. Assuming the normally ordered form of $\hat{f}$ exists,
one can write $\hat{f} = \sum^{+\infty}_{\vec{k}, \vec{l} = 0}
c_{\vec{k}\vec{l}} \hat{\vec{a}}^{\dagger \vec{k}} \hat{\vec{a}}^{\vec{l}}$,
where $\vec{k} = (k_1, \ldots, k_n)$, $\vec{l} = (l_1, \ldots, l_n)$ and
$\hat{\vec{a}}^{\dagger \vec{k}}$ ($\hat{\vec{a}}^{\vec{l}}$) means
$\hat{a}^{\dagger k_1}_1 \ldots \hat{a}^{\dagger k_n}_n$ ($\hat{a}^{l_1}_1
\ldots \hat{a}^{l_n}_n$). The mean value $\langle \hat{f}^\dagger \hat{f}
\rangle_A$ is a quadratic form with respect to the coefficients
$c_{\vec{k}\vec{l}}$,
\begin{equation}\label{eq:ff2}
    \langle \hat{f}^\dagger \hat{f} \rangle_A = \sum^{+\infty}_{\vec{k}, \vec{l}, \vec{p}, \vec{q} = 0}
    c^*_{\vec{k}\vec{l}} c_{\vec{p}\vec{q}} \langle \hat{\vec{a}}^{\dagger \vec{l}} \hat{\vec{a}}^{\vec{k}}
    \hat{\vec{a}}^{\dagger \vec{p}} \hat{\vec{a}}^{\vec{q}} \rangle_A.
\end{equation}
To write this quadratic form in the standard way, let us order the
multi-indices $(\vec{k}, \vec{l})$ and numerate them with a single number. The
exact order is unimportant; for convenience we use the following one: for any
two multi-indices $\vec{u} = (\vec{k}, \vec{l})$ and $\vec{v} = (\vec{p},
\vec{q})$
\begin{equation}\label{eq:gralex}
    \vec{u} < \vec{v} \leftrightarrow
    \begin{cases}
        |\vec{u}| < |\vec{v}| \ \text{or} \\
        |\vec{u}| = |\vec{v}| \ \text{and} \ \vec{u} <' \vec{v},
    \end{cases}
\end{equation}
where $|\vec{u}| = \sum_i k_i + \sum_i l_i$ and $\vec{u} <' \vec{v}$ means that the first nonzero difference $p_n-k_n,
q_n-l_n, \ldots, p_1-k_1, q_1-l_1$ is positive. Combining the pairs of $n$-dimensional multi-indices $(\vec{r}, \vec{s})$ into the single $2n$-dimensional ones as $(\vec{r}, \vec{s}) \to (s_1, r_1, \ldots, s_n, r_n)$, we see that the order \eqref{eq:gralex} becomes the graded antilexicographical order, which is defined for any
dimension $d$ as follows: for two $d$-dimensional indices $\vec{u} = (u_1, \ldots, u_d)$ and $\vec{v} = (v_1, \ldots,
v_d)$
\begin{equation}
    \vec{u} <^{\mathrm{gralex}} \vec{v} \leftrightarrow
    \begin{cases}
        |\vec{u}| < |\vec{v}| \ \text{or} \\
        |\vec{u}| = |\vec{v}| \ \text{and} \ \vec{u} <^{\mathrm{alex}} \vec{v},
    \end{cases}
\end{equation}
where $\vec{u} <^{\mathrm{alex}} \vec{v}$ means antilexicographical order,
i.e. the first nonzero difference $v_d-u_d, \ldots, v_1-u_1$ is positive. The
resulting ordered sequence of the moments starts as follows: $1, \langle
\hat{a}_1 \rangle, \langle \hat{a}^\dagger_1 \rangle, \ldots, \langle
\hat{a}_n \rangle, \langle \hat{a}^\dagger_n \rangle, \ldots$. The quadratic
form \eqref{eq:ff2} can now be written as $\langle \hat{f}^\dagger \hat{f}
\rangle_A = \sum^{+\infty}_{r, s = 0} M_{rs} c^*_r c_s$, where $M_{rs} =
\langle \hat{\vec{a}}^{\dagger \vec{l}} \hat{\vec{a}}^{\vec{k}}
\hat{\vec{a}}^{\dagger \vec{p}} \hat{\vec{a}}^{\vec{q}} \rangle_A$, $c_r =
c_{\vec{k}\vec{l}}$, $c_s = c_{\vec{p}\vec{q}}$ with $(\vec{k}, \vec{l})$ and
$(\vec{p}, \vec{q})$ being the $r$th and the $s$th indices, respectively, in
the ordered sequence of moments.

The nonnegativity of this quadratic form is equivalent to the nonnegativity of all its principal minors. That is, $\langle \hat{f}^\dagger \hat{f} \rangle_A \geqslant 0$, if and only if the conditions
\begin{equation}\label{eq:D}
    D_R =
    \begin{vmatrix}
        M_{r_1 r_1} & \ldots & M_{r_1 r_N} \\
        \hdotsfor{3} \\
        M_{r_N r_1} & \ldots & M_{r_N r_N}
    \end{vmatrix}
    \geqslant 0
\end{equation}
are satisfied for all $N = 1, 2, \ldots, \infty$ and for all $R = \{r_1, \ldots, r_N\}$ with $1 \leqslant r_1 < \ldots < r_N$. For $R = \{1, 2, \ldots, 2n+1\}$ the corresponding minor is denoted by $D_{2n+1}$ and it reads as
\begin{equation}\label{eq:D2n1}
    D_{2n+1} =
    \begin{vmatrix}
        1 & \langle \hat{a}_1 \rangle & \langle \hat{a}^\dagger_1 \rangle & \ldots & \langle \hat{a}^\dagger_n \rangle \\
        \langle \hat{a}^\dagger_1 \rangle & \langle \hat{a}^{\dagger}_1 \hat{a}_1 \rangle &
        \langle \hat{a}^{\dagger 2}_1 \rangle & \ldots & \langle \hat{a}^\dagger_1 \hat{a}^\dagger_n \rangle \\
        \ldots & \ldots & \ldots & \ldots & \ldots \\
        \langle \hat{a}_n \rangle & \langle \hat{a}_1 \hat{a}_n \rangle & \langle \hat{a}^\dagger_1 \hat{a}_n \rangle & \ldots & \langle \hat{a}^{\dagger}_n \hat{a}_n \rangle + 1
    \end{vmatrix}.
\end{equation}
For any nonnegative operator $\hat{A}$ the determinant $D_{2n+1}$ is nonnegative: the condition $D_{2n+1} \geqslant 0$
is necessary for nonnegativity of $\hat{A}$ (we have omitted the subscript $A$).

Now we can apply the conditions \eqref{eq:D} to the partially transposed density operator
$\hat{A} = \mathrm{PT}_I(\hat{\varrho})$: it is nonnegative defined if and only if the conditions
\begin{equation}\label{eq:D2}
    D_R^I =
    \begin{vmatrix}
        M^I_{r_1 r_1} & \ldots & M^I_{r_1 r_N} \\
        \hdotsfor{3} \\
        M^I_{r_N r_1} & \ldots & M^I_{r_N r_N}
    \end{vmatrix}
    \geqslant 0
\end{equation}
are satisfied for all $N$, where $M^I_{st}$ are the moments $M_{st}$ calculated on the partially transposed density
operator $\mathrm{PT}_I(\hat{\varrho})$. The moments $M_{st} = \langle \hat{\vec{a}}^{\dagger \vec{l}}
\hat{\vec{a}}^{\vec{k}} \hat{\vec{a}}^{\dagger \vec{p}} \hat{\vec{a}}^{\vec{q}} \rangle$ of any quantum state can be written as
\begin{equation}
    M_{st} = \left\langle \prod^n_{i=1} \hat{a}^{\dagger l_i}_i \hat{a}^{k_i}_i \hat{a}^{\dagger p_i}_i \hat{a}^{q_i}_i
    \right\rangle.
\end{equation}
The moments $M^I_{st}$ of the partially transposed density operator read as
\begin{equation}
    M^I_{st} = \left\langle \prod_{i \in I} \hat{a}^{\dagger q_i}_i \hat{a}^{p_i}_i \hat{a}^{\dagger k_i}_i \hat{a}^{l_i}_i
    \prod_{i \in \overline{I}} \hat{a}^{\dagger l_i}_i \hat{a}^{k_i}_i \hat{a}^{\dagger p_i}_i \hat{a}^{q_i}_i \right\rangle.
\end{equation}
The other way around, the conditions \eqref{eq:D2} can be formulated as
follows: if there are $N$ indices $1 \leqslant r_1 < \ldots < r_N$ and a
non-empty subset $I \subseteq \mathcal{N}_{n-1}$ such that $D_R^I < 0$, than
the state under consideration is entangled. Note that, for the case of $n=2$
and $I = \{2\}$ the condition $D^I_5<0$ is exactly the Simon condition
\cite{prl-84-2726, prl-95-230502}.

\begin{figure}
    \includegraphics{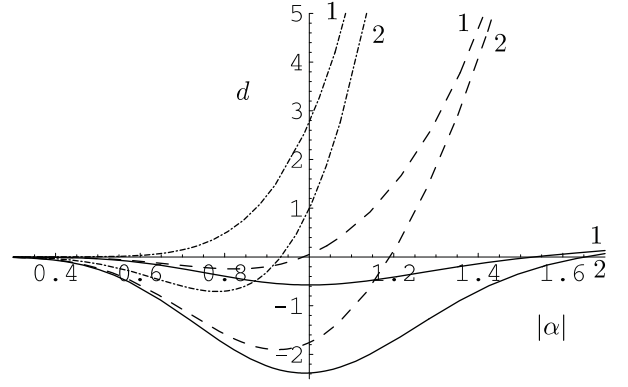}
\caption{The minors $d_1$ and $d_2$ are given for different noise levels: $\overline{n}=0$ (solid line), $\overline{n}=0.01$ (dashed line) and $\overline{n}=0.05$ (dot-dashed line).}\label{fig:1}
\end{figure}

As an example let us consider the mixed state
\begin{equation}
    \hat{\varrho}(\vec{\alpha}) = N(\vec{\alpha})\int P_{\overline{\vec{n}}}(\vec{\beta}, \vec{\alpha})
    |\psi(\vec{\beta})\rangle\langle\psi(\vec{\beta})| \, d^2\vec{\beta},
\end{equation}
where the pure state $|\psi(\vec{\beta})\rangle$ is defined by
\begin{equation}
    |\psi(\vec{\beta})\rangle = \sum^n_{i=1}|\beta_1, \ldots, \beta_{i-1}, -\beta_i, \beta_{i+1}, \ldots \beta_n\rangle,
\end{equation}
and $P_{\overline{\vec{n}}}(\vec{\beta}, \vec{\alpha})$ describes noise of mean photon number $\overline{\vec{n}}$,
\begin{equation}
    P_{\overline{\vec{n}}}(\vec{\beta}, \vec{\alpha}) = \prod^n_{i=1}P_{\overline{n}_i}(\beta_i, \alpha_i), \quad
    P_{\overline{n}}(\beta, \alpha) = \frac{1}{\pi \overline{n}} e^{-\frac{|\beta-\alpha|^2}{\overline{n}}}.
\end{equation}
This state is a noisy cv analog of the $W$-state. Here we consider the
four-mode case, $n=4$. For simplicity, we may start to consider the $2 \times
2$ principal minors $d_{\{13, 35\}} = d_{(1,2;3,4)}$, $d_{\{20, 33\}} = d_{(1, 3;2, 4)}$ and $d_{\{22, 31\}} = d_{(2, 3; 1, 4)}$, where $d_{(i,j;k,l)}$ is
defined via
\begin{equation}
\begin{split}
    d_{(i,j;k,l)} &=
    \begin{vmatrix}
        \langle \hat{a}^\dagger_i \hat{a}_i \hat{a}^\dagger_j \hat{a}_j \rangle &
        \langle \hat{a}^\dagger_i \hat{a}^\dagger_j \hat{a}_k \hat{a}_l \rangle \\
        \langle \hat{a}_i \hat{a}_j \hat{a}^\dagger_k \hat{a}^\dagger_l \rangle &
        \langle \hat{a}^\dagger_k \hat{a}_k \hat{a}^\dagger_l \hat{a}_l \rangle
    \end{vmatrix}.
\end{split}
\end{equation}
This principal minor corresponds to the operator $\hat{f} = c_1 \hat{a}_i \hat{a}_j + c_2 \hat{a}_k \hat{a}_l$ in the expression \eqref{eq:ff2}. The behavior of the partially transposed minors $d_1 = d^{\{1\}}_{\{13, 35\}} = d^{\{2\}}_{\{13, 35\}} = d^{\{3\}}_{\{13, 35\}} = d^{\{1, 2, 3\}}_{\{13, 35\}}$ and  $d_2 = d^{\{1, 2\}}_{\{13, 35\}} = d^{\{1, 3\}}_{\{20, 33\}} = d^{\{2, 3\}}_{\{22, 31\}}$ is shown in Fig. \ref{fig:1}. Note that these minors coincide only for our highly symmetric case ($\alpha_i = \alpha$ and $\overline{n}_i = \overline{n}$ for $i = 1, 2, 3, 4$). Figure \ref{fig:1} illustrates the behavior of the minors $d_1$ and $d_2$ as function of $|\alpha|$ for different noise levels. It shows that the maximal negativity can be found for the noiseless case ($\overline{n}=0$). One can see that for a range of $|\alpha|$ (which depends on the noise) the state under study is fully entangled, what can be demonstrated already by lower-order moments. For larger values of $|\alpha|$ or for a larger noise level lower-order moments do not show the negativity.

To certify full multipartite entanglement, one needs a systematic procedure to
deal with the data of a given quantum state to be analyzed. The data may be
calculated or measured moments, depending on whether a theoretical or an
experimentally measured quantum state is considered. Let us present in detail the algorithm for computing the $n$th multi-index in the graded antilexicographical order. First, for any $d$-dimensional index $\vec{u}$ let us find the next multi-index $\vec{v}$, which has
the number $n+1$, i.e. such an index $\vec{v}$ that $\vec{u}
<^{\mathrm{gralex}} \vec{v}$ but there are no other indices $\vec{w}$ with
$\vec{u} <^{\mathrm{gralex}} \vec{w} <^{\mathrm{gralex}} \vec{v}$. The index
$\vec{v}$, next to the given $\vec{u}$, can be computed with the following
algorithm: \\
\textbf{nextMultiindex($\vec{u}$)} \\
\textbf{input}: $\vec{u} = (u_1, \ldots, u_d)$ \\
\textbf{output}: minimal $\vec{v}$ such that $\vec{u} <^{\mathrm{gralex}}
\vec{v}$
\begin{enumerate}
\item \textit{find the minimal $i$ such that $u_i \not= 0$; if all $u_i=0$ set $i = d$}.
\item \textbf{if} $i=d$ \\
      \hspace*{3mm} $\vec{v} = (u_d+1, 0, \ldots, 0)$; \\
\textbf{else} \\
      \hspace*{3mm} $\vec{v} = (\underbrace{u_i-1, 0, \ldots, 0}_i, u_{i+1}+1, u_{i+2}, \ldots, u_d)$.
\end{enumerate}
The $n$th multi-index can be computed using this function repeatedly starting with the minimal multi-index $(0, \ldots, 0)$:
\begin{equation}
    \nthM(d, n) = \nextM^{[n-1]}(\underbrace{0, \ldots, 0}_d),
\end{equation}
where $f^{[0]}(\vec{u}) = \vec{u}$ and $f^{[n]}(\vec{u}) = f(f^{[n-1]}(\vec{u}))$ for all $n>0$. In the case of $d=2$ one can get the explicit expression for the $n$th two-dimensional index:
\begin{equation}
\begin{split}
    &\nthM(2, n) = \\
    &\left(\frac{(N+1)(N+2)}{2}-n, n-\frac{N(N+1)}{2}-1\right),
\end{split}
\end{equation}
where $N = \lceil \frac{\sqrt{8n+1}-3}{2} \rceil$, with $\lceil x \rceil$ being the smallest integer greater than or equal to $x$.

Let us comment on the realization of the measurements to verify full
multipartite entanglement. Our chosen example of a fully entangled quantum
state clearly shows the strong sensitivity of entanglement with respect to
noise effects. Any quantum-state reconstruction based on
homodyne detection with imperfect detectors includes Gaussian noise effects, for a review
see~\cite{welsch99}, that may prevent one from
verifying full entanglement. Therefore we have recently proposed the method of
balanced homodyne correlation measurements to overcome such
problems~\cite{prl-96-200403}. Upon properly balancing the setup, any kind
of minor to be considered in our method is simply proportional to a
corresponding power of the detection efficiency. Though different moments are proportional to different powers of the quantum efficiency, in the minors they are combined in such a way that the whole minor is proportional to one common power of the efficiency. Hence all the inequalities to be used to verify the nonpositivity of partial transpositions are only
multiplied by a positive factor, so that the method also works in the case of
imperfect detection.
 
In conclusion, we have applied the partial transposition approach to get insight into multipartite entanglement of mixed multi-mode cv quantum states. The method allows us to certify full entanglement of a broad class of multimode quantum states, also of non-Gaussian ones, which requires making use of higher-order moments. We also provide a \textit{Mathematica} package to automate	 the process of testing partial transpositions, given that the normally ordered moments are defined either analytically or as a result of measurements.

\end{document}